\documentclass[fleqn,twoside]{article}
\usepackage{amsmath}
\usepackage{espcrc2}

\usepackage{graphicx}
\usepackage[figuresright]{rotating}
\newcommand{\half}{{\frac{1}{2}}}
\newcommand{\mbf}[1]{\mathbf{#1}}

\title{Light-Front Holography, AdS/QCD, and Hadronic Phenomena}
\author{Stanley J. Brodsky\address{SLAC National Accelerator Laboratory, 
Stanford University, Stanford, CA 94309, USA} 
and
Guy F. de T\'eramond\address{Universidad de Costa Rica, San Jos\'e, Costa Rica}}
\begin{document}
\begin{abstract}
AdS/QCD, the correspondence between theories in a modified five-dimensional anti-de Sitter space and
confining field theories in physical space-time,  provides a remarkable 
semiclassical model for hadron physics.  
Light-front holography allows hadronic amplitudes in the
AdS fifth dimension to be mapped to frame-independent light-front
wavefunctions of hadrons in physical space-time, thus providing a
relativistic description of hadrons at the amplitude level. We
identify the AdS coordinate $z$ with an invariant light-front
coordinate $\zeta$ which separates the dynamics of quark and gluon binding from 
the kinematics of constituent spin and internal orbital angular momentum. The result is a single-variable
light-front Schr\"odinger equation with a confining potential which determines the eigenspectrum and the light-front wavefunctions of hadrons for general spin and orbital angular momentum. The mapping of
electromagnetic and gravitational form factors in AdS space to their
corresponding expressions in light-front theory confirms this
correspondence. Some novel features  of QCD are discussed, including
the consequences of confinement for quark and gluon condensates. 
The distinction
between static structure functions, such as the probability
distributions  computed from the square of the light-front
wavefunctions, versus dynamical structure functions which include the
effects of rescattering, is emphasized. A new method for computing
the hadronization of quark and gluon jets at the amplitude level, an event amplitude generator, is outlined.

\vspace{1pc}
\end{abstract}

\maketitle

\section{The Light-Front Hamiltonian Approach to QCD}

One of the most important theoretical tools in atomic physics is the
Schr\"odinger wavefunction, which describes the quantum-mechanical
structure of  an atomic system at the amplitude level. Light-front
wavefunctions (LFWFs) play a similar role in quantum chromodynamics, 
providing a fundamental description of the structure and
internal dynamics of hadrons in terms of their constituent quarks
and gluons. The LFWFs of bound states in QCD are
relativistic generalizations of the Schr\"odinger wavefunctions of
atomic physics, but they are determined at fixed light-cone time
$\tau  = t +z/c$ -- the ``front form'' introduced by
Dirac~\cite{Dirac:1949cp} -- rather than at fixed ordinary time $t.$

When a flash from a camera illuminates a scene, each object is illuminated along the light-front of the flash; i.e., at a given $\tau$.  Similarly, when a sample is illuminated by an x-ray source, each element of the target is struck at a given $\tau.$  In contrast, setting the initial condition using conventional instant time $t$ requires simultaneous scattering of photons on each constituent. 
Thus it is natural to set boundary conditions at fixed $\tau$ and then evolve the system using the light-front (LF) Hamiltonian 
$P^-  \!= \!  P^0-P^3 = i {d/d \tau}.$  The invariant Hamiltonian $H_{LF} = P^+ P^- \! - P^2_\perp$ then has eigenvalues $\mathcal{M}^2$ where $\mathcal{M}$ is the physical mass.   Its eigenfunctions are the light-front eigenstates whose Fock state projections define the light-front wavefunctions.  Given the LF Fock state wavefunctions 
$\psi^H_n(x_i, \mbf{k}_{\perp i}, \lambda_i),$  
where $x_i \! =\! k^+/P^+$,  $\sum_{i=1}^n x_i  \! = \! 1, ~ \sum_{i=1}^n \mbf{k}_{\perp i}  \! = \! 0$, one
can immediately compute observables such as hadronic form factors (overlaps of LFWFs), structure
functions (squares of LFWFS), as well as the generalized parton distributions and
distribution amplitudes which underly hard exclusive reactions. 

A remarkable feature of LFWFs is the fact that they are frame
independent; i.e., the form of the LFWF is independent of the
hadron's total momentum $P^+ = P^0 + P^3$ and $\mbf{P}_\perp.$
The simplicity of Lorentz boosts of LFWFs contrasts dramatically with the complexity of the boost of wavefunctions defined at fixed time $t.$~\cite{Brodsky:1968ea}  
Light-front quantization is thus the ideal framework to describe the
structure of hadrons in terms of their quark and gluon degrees of freedom.  The
constituent spin and orbital angular momentum properties of the
hadrons are also encoded in the LFWFs.  
The total  angular momentum projection~\cite{Brodsky:2000ii} 
$J^z = \sum_{i=1}^n  S^z_i + \sum_{i=1}^{n-1} L^z_i$ 
is conserved Fock-state by Fock-state and by every interaction in the LF Hamiltonian.
Other advantageous features of light-front quantization include:

\begin{itemize}

\item
The simple structure of the light-front vacuum allows an unambiguous
definition of the partonic content of a hadron in QCD.  The chiral and gluonic condensates are properties of the higher Fock states~\cite{Casher:1974xd,Brodsky:2009zd}, rather than the vacuum.  In the case of the Higgs model, the effect of the usual Higgs vacuum expectation value is replaced by a constant $k^+=0$ zero mode field.~\cite{Srivastava:2002mw}

\item 
If one quantizes QCD in the physical light-cone gauge (LCG) $A^+ =0$, then gluons only have physical angular momentum projections $S^z= \pm 1$. The orbital angular momenta of quarks and gluons are defined unambiguously, and there are no ghosts.  

\item
The gauge-invariant distribution amplitude $\phi(x,Q)$  is the integral of the valence LFWF in LCG integrated over the internal transverse momentum $k^2_\perp < Q^2$ because the Wilson line is trivial in this gauge. It is also possible to quantize QCD in  Feynman gauge in the light front.~\cite{Srivastava:1999gi}

\item
LF Hamiltonian perturbation theory provides a simple method for deriving analytic forms for the analog of Parke-Taylor amplitudes~\cite{Motyka:2009gi} where each particle spin $S^z$ is quantized in the LF $z$ direction.  The gluonic $g^6$ amplitude  $T(-1 -1 \to +1 +1 +1 +1 +1 +1)$  requires $\Delta L^z =8;$ it thus must vanish at tree level since each three-gluon vertex has  $\Delta L^z = \pm 1.$ However, the order $g^8$ one-loop amplitude can be nonzero.

\item
Amplitudes in light-front perturbation theory are automatically renormalized using the ``alternate denominator"  subtraction method~\cite{Brodsky:1973kb}.  The application to QED has been checked at one and two loops.~\cite{Brodsky:1973kb}

\item 
One can easily show using LF quantization that the anomalous gravitomagnetic moment $B(0)$  of a nucleon, as  defined from the spin flip matrix element of the gravitational current, vanishes Fock-state by Fock state~\cite{Brodsky:2000ii}, as required by the equivalence principle.~\cite{Teryaev:1999su}

\item
LFWFs obey the cluster decomposition theorem, providing the only proof of this theorem for relativistic bound states.~\cite{Brodsky:1985gs}

\item
The LF Hamiltonian can be diagonalized using the discretized light-cone quantization (DLCQ) method.~\cite{Pauli:1985ps} This nonperturbative method is particularly useful for solving low-dimension quantum field theories such as 
QCD$(1+1).$~\cite{Hornbostel:1988fb}

\item 
LF quantization provides a distinction between static  (square of LFWFs) distributions versus non-universal dynamic structure functions,  such as the Sivers single-spin correlation and diffractive deep inelastic scattering which involve final state interactions.  The origin of nuclear shadowing and process independent anti-shadowing also becomes explicit.   This is discussed further in Sec.
\ref{rescat}.

\item 
LF quantization provides a simple method to implement jet hadronization at the amplitude level.  This is discussed in Sec. 
\ref{hadronization}.

\item 
The instantaneous fermion interaction in LF  quantization provides a simple derivation of the $J=0$
fixed pole contribution to deeply virtual Compton scattering.~\cite{Brodsky:2009bp}

\item
Unlike instant time quantization, the Hamiltonian equation of motion in the LF is frame independent. This makes a direct connection of QCD with AdS/CFT methods possible.~\cite{deTeramond:2008ht}

\end{itemize}

\section{Light-Front Holography}

A key step in the analysis of an atomic system such as positronium
is the introduction of the spherical coordinates $r, \theta, \phi$
which  separates the dynamics of Coulomb binding from the
kinematical effects of the quantized orbital angular momentum $L$.
The essential dynamics of the atom is specified by the radial
Schr\"odinger equation whose eigensolutions $\psi_{n,L}(r)$
determine the bound-state wavefunction and eigenspectrum. In our recent 
work, we have shown that there is an analogous invariant
light-front coordinate $\zeta$ which allows one to separate the
essential dynamics of quark and gluon binding from the kinematical
physics of constituent spin and internal orbital angular momentum.
The result is a single-variable LF Schr\"odinger equation for QCD
which determines the eigenspectrum and the light-front wavefunctions
of hadrons for general spin and orbital angular momentum.~\cite{deTeramond:2008ht}  If one further chooses  the constituent rest frame (CRF)~\cite{Danielewicz:1978mk,Karmanov:1979if,Glazek:1983ba}  where $\sum^n_{i=1} \mbf{k}_i \! = \! 0$, then the kinetic energy in the LFWF displays the usual 3-dimensional rotational invariance. Note that if the binding energy is nonzero, $P^z \ne 0,$ in this frame.

Light-Front Holography can be derived by observing the correspondence between matrix elements obtained in AdS/CFT with the corresponding formula using the LF 
representation.~\cite{Brodsky:2006uqa}  The light-front electromagnetic form factor in impact 
space~\cite{Brodsky:2006uqa,Brodsky:2007hb,Soper:1976jc} can be written as a sum of overlap of light-front wave functions of the $j = 1,2, \cdots, n-1$ spectator
constituents:
\begin{multline} \label{eq:FFb}
F(q^2) =  \sum_n  \prod_{j=1}^{n-1}\int d x_j d^2 \mbf{b}_{\perp j}   \sum_q e_q  \\ \times
            \exp \! {\Bigl(i \mbf{q}_\perp \! \cdot \sum_{j=1}^{n-1} x_j \mbf{b}_{\perp j}\Bigr)} 
 \left\vert  \psi_n(x_j, \mbf{b}_{\perp j})\right\vert^2,
\end{multline}
where the normalization is defined by
\vspace{-4pt}
\begin{equation}  \label{eq:Normb}
\sum_n  \prod_{j=1}^{n-1} \int d x_j d^2 \mathbf{b}_{\perp j}
\vert \psi_{n/H}(x_j, \mathbf{b}_{\perp j})\vert^2 = 1.
\end{equation}

The formula  is exact if the sum is over all Fock states $n$.
For definiteness we shall consider a two-quark $\pi^+$  valence Fock state 
$\vert u \bar d\rangle$ with charges $e_u = \frac{2}{3}$ and $e_{\bar d} = \frac{1}{3}$.
For $n=2$, there are two terms which contribute to the $q$-sum in (\ref{eq:FFb}). 
Exchanging $x \leftrightarrow 1 \! - \! x$ in the second integral  we find 
\begin{multline}  \label{eq:PiFFb}
 F_{\pi^+}(q^2)  =  2 \pi \int_0^1 \! \frac{dx}{x(1-x)}  \int \zeta d \zeta \\ \times
J_0 \! \left(\! \zeta q \sqrt{\frac{1-x}{x}}\right) 
\left\vert \psi_{u \bar d/ \pi}\!(x,\zeta)\right\vert^2,
\end{multline}
where $\zeta^2 =  x(1  -  x) \mathbf{b}_\perp^2$ and $F_{\pi^+}(q\!=\!0)=1$.

We now compare this result with the electromagnetic form-factor in AdS space~\cite{Polchinski:2002jw}:
\begin{equation} 
F(Q^2) = R^3 \int \frac{dz}{z^3} \, J(Q^2, z) \vert \Phi(z) \vert^2,
\label{eq:FFAdS}
\end{equation}
where $J(Q^2, z) = z Q K_1(z Q)$.
Using the integral representation of $J(Q^2,z)$
\begin{equation} \label{eq:intJ}
J(Q^2, z) = \int_0^1 \! dx \, J_0\negthinspace \left(\negthinspace\zeta Q
\sqrt{\frac{1-x}{x}}\right) ,
\end{equation} we write the AdS electromagnetic form-factor as
\begin{equation} 
F(Q^2)  =    R^3 \! \int_0^1 \! dx  \! \int \frac{dz}{z^3} \, 
J_0\!\left(\!z Q\sqrt{\frac{1-x}{x}}\right) \left \vert\Phi(z) \right\vert^2 .
\label{eq:AdSFx}
\end{equation}
Comparing with the light-front QCD  form factor (\ref{eq:PiFFb}) for arbitrary  values of $Q$~\cite{Brodsky:2006uqa}
\begin{equation} \label{eq:Phipsi} 
\vert \psi(x,\zeta)\vert^2 = 
\frac{R^3}{2 \pi} \, x(1-x)
\frac{\vert \Phi(\zeta)\vert^2}{\zeta^4}, 
\end{equation}
where we identify the transverse LF variable $\zeta$, $0 \leq \zeta \leq \Lambda_{\rm QCD}$,
with the holographic variable $z$.

Matrix elements of the energy-momentum tensor $\Theta^{\mu \nu} $ which define the gravitational form factors play an important role in hadron physics.  Since one can define $\Theta^{\mu \nu}$ for each parton, one can identify the momentum fraction and  contribution to the orbital angular momentum of each quark flavor and gluon of a hadron. For example, the spin-flip form factor $B(q^2)$, which is the analog of the Pauli form factor $F_2(Q^2)$ of a nucleon, provides a  measure of the orbital angular momentum carried by each quark and gluon constituent of a hadron at $q^2=0.$   Similarly,  the spin-conserving form factor $A(q^2)$, the analog of the Dirac form factor $F_1(q^2)$, allows one to measure the momentum  fractions carried by each constituent.
This is the underlying physics of Ji's sum rule~\cite{Ji:1996ek}:
$\langle J^z\rangle = \half [ A(0) + B(0)] $,  which has prompted much of the current interest in 
the generalized parton distributions (GPDs)  measured in deeply
virtual Compton scattering.  An important constraint is $B(0) = \sum_i B_i(0) = 0$;  i.e.  the anomalous gravitomagnetic moment of a hadron vanishes when summed over all the constituents $i$. This was originally derived from the equivalence principle of gravity~\cite{Teryaev:1999su}.  The explicit verification of these relations, Fock state by Fock state, can be obtained in the LF quantization of QCD in  light-cone 
gauge~\cite{Brodsky:2000ii}.  Physically $B(0) =0$ corresponds to the fact that the sum of the $n$ orbital angular momenta $L$ in an $n$-parton Fock state must vanish since there are only $n-1$ independent orbital angular momenta.

The LF expression for the helicity-conserving gravitational form factor in impact space
is~\cite{Brodsky:2008pf}
\begin{multline} \label{eq:Ab}
A(q^2) =  \sum_n  \prod_{j=1}^{n-1}\int d x_j d^2 \mbf{b}_{\perp j}  \sum_f x_f \cr \times
\exp \! {\Bigl(i \mbf{q}_\perp \! \cdot \sum_{j=1}^{n-1} x_j \mbf{b}_{\perp j}\Bigr)} 
\left\vert  \psi_n(x_j, \mbf{b}_{\perp j})\right\vert^2,
\end{multline}
which includes the contribution of each struck parton with longitudinal momentum $x_f$
and corresponds to a change of transverse momentum $x_j \mbf{q}$ for
each of the $j = 1, 2, \cdots, n-1$ spectators. 
For $n=2$, there are two terms which contribute to the $f$-sum in  (\ref{eq:Ab}). 
Exchanging $x \leftrightarrow 1-x$ in the second integral we find 
\begin{multline} \label{eq:PiGFFb}
A_{\pi}(q^2) =  4 \pi \int_0^1 \frac{dx}{(1-x)}  \int \zeta d \zeta \\ \times
J_0 \! \left(\! \zeta q \sqrt{\frac{1-x}{x}}\right) 
\left\vert \psi_{q \bar q/ \pi}\!(x,\zeta)\right\vert^2,
\end{multline}
where $\zeta^2 =  x(1-x) \mathbf{b}_\perp^2$ and  $A_{\pi}(0) = 1$.
 We now consider the expression for the hadronic gravitational form factor in AdS space~\cite{Abidin:2008ku}
\begin{equation} 
A_\pi(Q^2)  =  R^3 \! \! \int \frac{dz}{z^3} \, H(Q^2, z) \left\vert\Phi_\pi(z) \right\vert^2,
\end{equation}
where $H(Q^2, z) = \half  Q^2 z^2  K_2(z Q)$ and $A(0) = 1$.
Using the integral representation of $H(Q^2,z)$
\begin{equation} \label{eq:intHz}
H(Q^2, z) =  2  \int_0^1\!  x \, dx \, J_0\!\left(\!z Q\sqrt{\frac{1-x}{x}}\right) ,
\end{equation}
we can write the AdS gravitational form factor 
\begin{multline} 
A(Q^2)  =  2  R^3 \! \int_0^1 \! x \, dx  \! \int \frac{dz}{z^3}  \\ \times
J_0\!\left(\!z Q\sqrt{\frac{1-x}{x}}\right) \left \vert\Phi(z) \right\vert^2 .
\label{eq:AdSAx}
\end{multline}
Comparing with the QCD  gravitational form factor (\ref{eq:PiGFFb}) we find an  identical  relation between the LF wave function $\psi(x,\zeta)$ and the AdS wavefunction $\Phi(z)$
given in Eq. (\ref{eq:Phipsi}) which was obtained from the mapping of the pion electromagnetic transition amplitude.

One can also derive light-front holography using a first semiclassical approximation  to transform the fixed 
light-front time bound-state Hamiltonian equation of motion in QCD  to  a corresponding wave equation in AdS 
space.~\cite{deTeramond:2008ht} To this end we
 compute the invariant hadronic mass $\mathcal{M}^2$ from the hadronic matrix element
\begin{equation}
\langle \psi_H(P') \vert H_{LF}\vert\psi_H(P) \rangle  = 
\mathcal{M}_H^2  \langle \psi_H(P' ) \vert\psi_H(P) \rangle,
\end{equation}
expanding the initial and final hadronic states in terms of its Fock components. We use the 
frame $P = \big(P^+, M^2/P^+, \vec{0}_\perp \big)$ where $H_{LF} =  P^+ P^-$.
The LF expression for $\mathcal{M}^2$ In impact space is
\begin{multline}   
 \mathcal{M}_H^2  =  \sum_n  \prod_{j=1}^{n-1} \int d x_j \, d^2 \mbf{b}_{\perp j} \,
\psi_n^*(x_j, \mbf{b}_{\perp j}) \\
\times  \sum_q   \left(\frac{ \mbf{- \nabla}_{ \mbf{b}_{\perp q}}^2  \! + m_q^2 }{x_q} \right) 
 \psi_n(x_j, \mbf{b}_{\perp j}) \\
  + {\rm (interactions)} , \label{eq:Mb}
 \end{multline}
plus similar terms for antiquarks and gluons ($m_g = 0)$. 

To simplify the discussion we will consider a two-parton hadronic bound state.  In the limit of zero quark masses
$m_q \to 0$
\begin{multline}  \label{eq:Mb}
\mathcal{M}^2  =  \int_0^1 \! \frac{d x}{x(1-x)} \int  \! d^2 \mbf{b}_\perp  \,
  \psi^*(x, \mbf{b}_\perp) \\ \times
  \left( - \mbf{\nabla}_{ {\mbf{b}}_{\perp}}^2\right)
  \psi(x, \mbf{b}_\perp) +   {\rm (interactions)}.
 \end{multline}

 The functional dependence  for a given Fock state is
given in terms of the invariant mass
\begin{equation}
 \mathcal{M}_n^2  = \Big( \sum_{a=1}^n k_a^\mu\Big)^2 = \sum_a \frac{\mbf{k}_{\perp a}^2 +  m_a^2}{x_a}
 \to \frac{\mbf{k}_\perp^2}{x(1-x)} \,,
 \end{equation}
 the measure of the 
 off-energy shell of the bound state, $\mathcal{M}^2 \! - \! \mathcal{M}_n^2$.
 Similarly in impact space the relevant variable for a two-parton state is  $\zeta^2= x(1-x)\mbf{b}_\perp^2$.
Thus, to first approximation  LF dynamics  depend only on the boost invariant variable
$\mathcal{M}_n$ or $\zeta,$
and hadronic properties are encoded in the hadronic mode $\phi(\zeta)$ from the relation
\begin{equation} \label{eq:psiphi}
\psi(x,\zeta, \varphi) = e^{i M \varphi} X(x) \frac{\phi(\zeta)}{\sqrt{2 \pi \zeta}} ,
\end{equation}
thus factoring out the angular dependence $\varphi$ and the longitudinal, $X(x)$, and transverse mode $\phi(\zeta)$
with normalization $ \langle\phi\vert\phi\rangle = \int \! d \zeta \,
 \vert \langle \zeta \vert \phi\rangle\vert^2 = 1$. The mapping  of transition matrix elements
 for arbitrary values of the momentum transfer~\cite{Brodsky:2006uqa,Brodsky:2007hb,Brodsky:2008pf} 
 gives $X(x) = \sqrt{x(1-x)}$.

We can write the Laplacian operator in (\ref{eq:Mb}) in circular cylindrical coordinates $(\zeta, \varphi)$
and factor out the angular dependence of the
modes in terms of the $SO(2)$ Casimir representation $L^2$ of orbital angular momentum in the
transverse plane. Using  (\ref{eq:psiphi}) we find~\cite{deTeramond:2008ht}
\begin{multline} \label{eq:KV}  
\mathcal{M}^2   =  \int \! d\zeta \, \phi^*(\zeta) \sqrt{\zeta}
\left( -\frac{d^2}{d\zeta^2} -\frac{1}{\zeta} \frac{d}{d\zeta}
+ \frac{L^2}{\zeta^2}\right)
\frac{\phi(\zeta)}{\sqrt{\zeta}}   \\
+ \int \! d\zeta \, \phi^*(\zeta) \, U(\zeta)  \, \phi(\zeta) ,
\end{multline}
where all the complexity of the interaction terms in the QCD Lagrangian is summed up in the effective potential $U(\zeta)$.
The light-front eigenvalue equation $H_{LF} \vert \phi \rangle = \mathcal{M}^2 \vert \phi \rangle$
is thus a LF wave equation for $\phi$
\begin{equation} \label{eq:QCDLFWE}
\left(-\frac{d^2}{d\zeta^2}
- \frac{1 - 4L^2}{4\zeta^2} + U(\zeta) \right) 
\phi(\zeta) = \mathcal{M}^2 \phi(\zeta),
\end{equation}
an effective single-variable light-front Schr\"odinger equation which is
relativistic, covariant and analytically tractable. Using (\ref{eq:Mb}) one can readily
generalize the equations to allow for the kinetic energy of massive
quarks~\cite{Brodsky:2008pg}.  In this case, however,
the longitudinal mode $X(x)$ does not decouple from the effective LF bound-state equations.

In the hard-wall model one has $U(z)=0$; confinement is introduced by requiring the wavefunction to vanish at $z=z_0 \equiv 1/\Lambda_{\rm QCD}.$~\cite{Polchinski:2001tt}
In the case of the soft-wall model,~\cite{Karch:2006pv}  the potential arises from a ``dilaton'' modification of the AdS metric; it  has the form of a harmonic oscillator  $ U(z) = \kappa^4 z^2 + 2 \kappa^2(L+S-1).$

\begin{figure}[!]
\includegraphics[width=7.5cm]{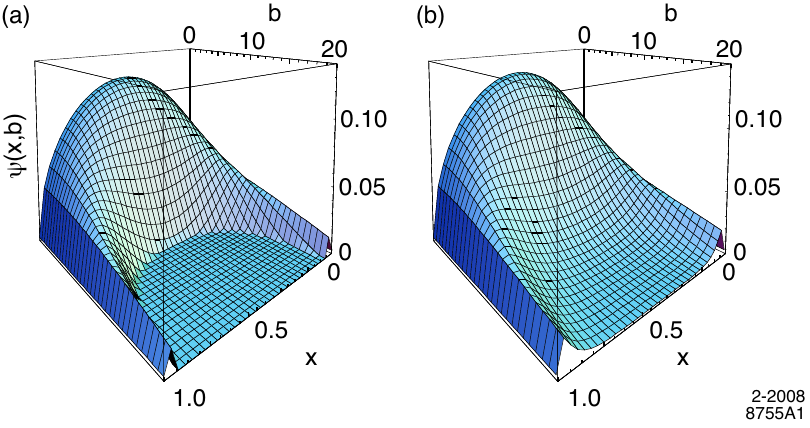}
 \caption{Pion light-front wavefunction $\psi_\pi(x, \mbf{b}_\perp$) for the  AdS/QCD (a) hard-wall ($\Lambda_{QCD} = 0.32$ GeV) and (b) soft-wall  ( $\kappa = 0.375$ GeV)  models.}
\label{LFWF}  
\end{figure}

The resulting mass spectra  for mesons  at zero quark mass is
${\cal M}^2 = 4 \kappa^2 (n + L +S/2)$
in the soft-wall model discussed here.
The spectral predictions for both light meson and baryon states for the hard and soft-wall holographic models discussed here are compared with experimental data in~\cite{Brodsky:2006uqa,Brodsky:2008pg,deTeramond:2005su,deTeramond:2009xx}.
The corresponding wavefunctions (see Fig.  \ref{LFWF})
display confinement at large interquark
separation and conformal symmetry at short distances, reproducing dimensional counting rules for hard exclusive amplitudes.

\section {Vacuum Effects and Light-Front Quantization}

The LF vacuum is remarkably simple in light-cone quantization because of the restriction $k^+ \ge 0.$   For example in QED,  vacuum graphs such as $e^+ e^- \gamma $  associated with the zero-point energy do not arise. In the Higgs theory, the usual Higgs vacuum expectation value is replaced with a $k^+=0$ zero mode~\cite{Srivastava:2002mw}; however, the resulting phenomenology is identical to the standard analysis.

Hadronic condensates play an important role in quantum chromodynamics (QCD).
Conventionally, these condensates are considered to be properties
of the QCD vacuum and hence to be constant throughout spacetime.
Recently a new perspective on the nature of QCD
condensates $\langle \bar q q \rangle$ and $\langle
G_{\mu\nu}G^{\mu\nu}\rangle$, particularly where they have spatial and temporal
support,
has been presented.~\cite{Brodsky:2009zd,Brodsky:2008be,Brodsky:2008xm,Brodsky:2008xu} 
Their spatial support is restricted to the interior
of hadrons, since these condensates arise due to the interactions of quarks and
gluons which are confined within hadrons. For example, consider a meson consisting of a light quark $q$ bound to a heavy
antiquark, such as a $B$ meson.  One can analyze the propagation of the light
$q$ in the background field of the heavy $\bar b$ quark.  Solving the
Dyson-Schwinger equation for the light quark one obtains a nonzero dynamical
mass and, via the connection mentioned above, hence a nonzero value of the
condensate $\langle \bar q q \rangle$.  But this is not a true vacuum
expectation value; instead, it is the matrix element of the operator $\bar q q$
in the background field of the $\bar b$ quark.  The change in the (dynamical)
mass of the light quark in this bound state is somewhat reminiscent of the
energy shift of an electron in the Lamb shift, in that both are consequences of
the fermion being in a bound state rather than propagating freely.
Similarly, it is important to use the equations of motion for confined quarks
and gluon fields when analyzing current correlators in QCD, not free
propagators, as has often been done in traditional analyses of operator
products.  Since after a $q \bar q$ pair is created, the distance between the
quark and antiquark cannot get arbitrarily great, one cannot create a quark
condensate which has uniform extent throughout the universe.  The $45$ orders of magnitude conflict of QCD with the observed value of the cosmological condensate is thus removed~\cite{Brodsky:2008xu}.
A new perspective on the nature of quark and gluon condensates in
quantum chromodynamics is thus obtained:~\cite{Brodsky:2008be,Brodsky:2008xm,Brodsky:2008xu}  the spatial support of QCD condensates
is restricted to the interior of hadrons, since they arise due to the
interactions of confined quarks and gluons.  In the LF theory, the condensate physics is replaced by the dynamics of higher non-valence Fock states as shown by Casher and Susskind.~\cite{Casher:1974xd}  In particular, chiral symmetry is broken in a limited domain of size $1/ m_\pi$,  in analogy to the limited physical extent of superconductor phases.  This novel description  of chiral symmetry breaking  in terms of ``in-hadron condensates"  has also been observed in Bethe-Salpeter studies.~\cite{Maris:1997hd,Maris:1997tm}
This picture explains the
results of recent studies~\cite{Ioffe:2002be,Davier:2007ym,Davier:2008sk} which find no significant signal for the vacuum gluon
condensate.

AdS/QCD also provides  a description of chiral symmetry breaking by
using the propagation of a scalar field $X(z)$
to represent the dynamical running quark mass. The AdS
solution has the form~\cite{Erlich:2005qh,DaRold:2005zs} $X(z) = a_1 z+ a_2 z^3$, where $a_1$ is
proportional to the current-quark mass. The coefficient $a_2$ scales as
$\Lambda^3_{QCD}$ and is the analog of $\langle \bar q q \rangle$; however,
since the quark is a color nonsinglet, the propagation of $X(z),$ and thus the
domain of the quark condensate, is limited to the region of color confinement. 
Furthermore the effect of the $a_2$ term
varies within the hadron, as characteristic of an in-hadron condensate. 
The AdS/QCD picture of condensates with spatial support restricted to hadrons
is also in general agreement with results from chiral bag 
models~\cite{Chodos:1975ix,Brown:1979ui,Hosaka:1996ee},
which modify the original MIT bag by coupling a pion field to the surface of
the bag in a chirally invariant manner.

\section{Hadronization at the Amplitude Level \label{hadronization}}

\begin{figure}[!]
\includegraphics[width=7.5cm]{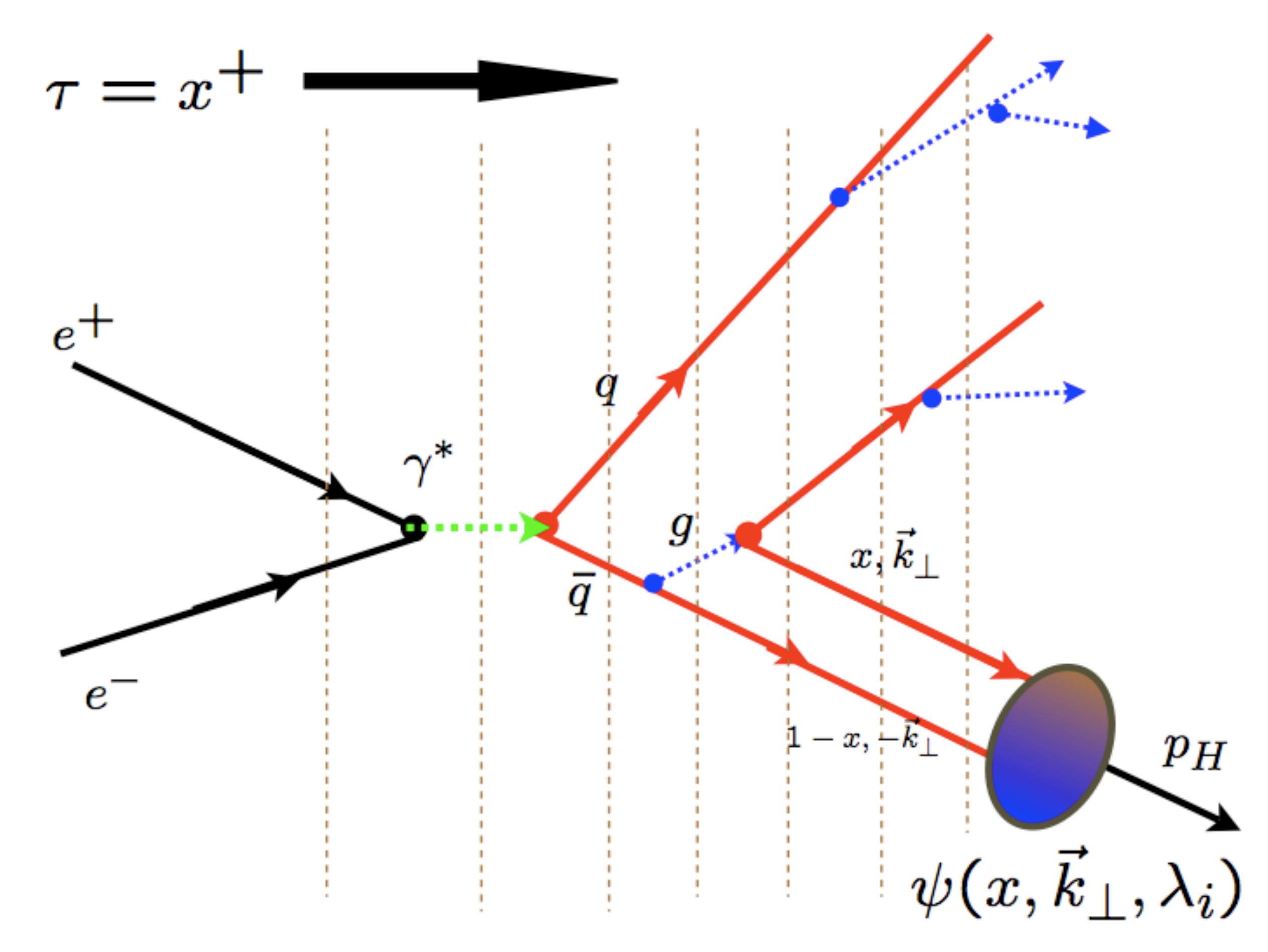}
  \caption{Illustration of an event amplitude generator for $e^+ e^- \to \gamma^* \to X$ for 
  hadronization processes at the amplitude level. Capture occurs if
  $\zeta^2 = x(1-x) \mbf{b}_\perp^2 > 1/ \Lambda_{\rm QCD}^2$
   in the AdS/QCD hard-wall model of confinement;  i.e., if
  $\mathcal{M}^2 = \frac{\mbf{k}_\perp^2}{x(1-x)} < \Lambda_{\rm QCD}^2$.}
\label{hadroniz}  
\end{figure}

The conversion of quark and gluon partons is usually discussed in terms  of on-shell hard-scattering cross sections convoluted with {\it ad hoc} probability distributions. 
The LF Hamiltonian formulation of quantum field theory provides a natural formalism to compute 
hadronization at the amplitude level.~\cite{Brodsky:2008tk}  In this case one uses light-front time-ordered perturbation theory for the QCD light-front Hamiltonian to generate the off-shell  quark and gluon T-matrix helicity amplitude  using the LF generalization of the Lippmann-Schwinger formalism:
\begin{multline}
T ^{LF}= 
{H^{LF}_I }  \\ + 
{H^{LF}_I }{1 \over {\cal M}^2_{\rm Initial} - {\cal M}^2_{\rm intermediate} + i \epsilon} {H^{LF}_I }  
+ \cdots 
\end{multline}
Here   ${\cal M}^2_{\rm intermediate} \!  = \! \sum^N_{i=1} {(\mbf{k}^2_{\perp i} + m^2_i )/x_i}$ is the invariant mass squared of the intermediate state and ${H^{LF}_I }$ is the set of interactions of the QCD LF Hamiltonian in the ghost-free light-cone gauge~\cite{Brodsky:1997de}.
The $T^{LF}$ matrix element is
evaluated between the out and in eigenstates of $H^{QCD}_{LF}$.   The event amplitude generator is illustrated for $e^+ e^- \to \gamma^* \to X$ in Fig. \ref{hadroniz}.

The LFWFS of AdS/QCD can be used as the interpolating amplitudes between the off-shell quark and gluons and the bound-state hadrons.
Specifically,
if at any stage a set of  color-singlet partons has  light-front kinetic energy 
$\sum_i {\mbf{k}^2_{\perp i}/ x_i} \!  < \! \Lambda^2_{\rm QCD}$, then one coalesces the virtual partons into a hadron state using the AdS/QCD LFWFs.   This provides a specific scheme for determining the factorization scale which  matches perturbative and nonperturbative physics.

This scheme has a number of  important computational advantages:

(a) Since propagation in LF Hamiltonian theory only proceeds as $\tau$ increases, all particles  propagate as forward-moving partons with $k^+_i \ge 0$.  There are thus relatively few contributing
 $\tau$-ordered diagrams.

(b) The computer implementation can be highly efficient: an amplitude of order $g^n$ for a given process only needs to be computed once.  In fact, each non-interacting cluster within $T^{LF}$ has a numerator which is process independent; only the LF denominators depend on the context of the process.  This method has recently been used by   L.~Motyka and A.~M.~Stasto~\cite{Motyka:2009gi}
to compute gluonic scattering amplitudes in QCD.

(c) Each amplitude can be renormalized using the ``alternate denominator'' counterterm method~\cite{Brodsky:1973kb}, rendering all amplitudes UV finite.

(d) The renormalization scale in a given renormalization scheme  can be determined for each skeleton graph even if there are multiple physical scales.

(e) The $T^{LF}$ matrix computation allows for the effects of initial and final state interactions of the active and spectator partons. This allows for leading-twist phenomena such as diffractive DIS, the Sivers spin asymmetry and the breakdown of the PQCD Lam-Tung relation in Drell-Yan processes.

(f)  ERBL and DGLAP evolution are naturally incorporated, including the quenching of  DGLAP evolution  at large $x_i$ where the partons are far off-shell.

(g) Color confinement can be incorporated at every stage by limiting the maximum wavelength of the propagating quark and gluons.

(h) This method retains the quantum mechanical information in hadronic production amplitudes which underlie Bose-Einstein correlations and other aspects of the spin-statistics theorem.
Thus Einstein-Podolsky-Rosen QM correlations are maintained even between far-separated hadrons and  clusters.

A similar off-shell T-matrix approach was used to predict antihydrogen formation from virtual positron--antiproton states produced in $\bar p A$ 
collisions~\cite{Munger:1993kq}.

\section{Dynamical Effects of Rescattering \label{rescat}}

Initial- and
final-state rescattering, neglected in the parton model, have a profound effect in QCD hard-scattering reactions,
predicting single-spin asymmetries~\cite{Brodsky:2002cx,Collins:2002kn}, diffractive deep lepton-hadron inelastic scattering~\cite{Brodsky:2002ue}, the breakdown of
the Lam Tung relation in Drell-Yan reactions~\cite{Boer:2002ju}, nor nuclear shadowing and non-universal 
antishadowing~\cite{Brodsky:2004qa}---leading-twist physics which is not incorporated in
the light-front wavefunctions of the target computed in isolation. 
It is thus important to distinguish~\cite{Brodsky:2008xe} ``static'' or ``stationary'' structure functions which are computed directly from the LFWFs of the target  from the ``dynamic'' empirical structure functions which take into account rescattering of the struck quark.   Since they derive from the LF eigenfunctions of the target hadron, the static structure functions have a probabilistic interpretation.  The wavefunction of a stable eigenstate is real; thus the static structure functions cannot describe diffractive deep inelastic scattering nor the single-spin asymmetries since such phenomena involves the complex phase structure of the $\gamma^* p $ amplitude.  
One can augment the light-front wavefunctions with a gauge link corresponding to an external field
created by the virtual photon $q \bar q$ pair
current~\cite{Belitsky:2002sm,Collins:2004nx}, but such a gauge link is
process dependent~\cite{Collins:2002kn}, so the resulting augmented
wavefunctions are not universal~\cite{Brodsky:2002ue,Belitsky:2002sm,Collins:2003fm}.  

It should be emphasized
that the shadowing of nuclear structure functions is due to the
destructive interference between multi-nucleon amplitudes involving
diffractive DIS and on-shell intermediate states with a complex
phase.  The physics of rescattering and shadowing is thus not
included in the nuclear light-front wavefunctions, and a
probabilistic interpretation of the nuclear DIS cross section is
precluded. 
In addition, one finds that antishadowing in deep inelastic lepton-nucleus scattering is is not universal~\cite{Brodsky:2004qa},
but depends on the flavor of each quark and antiquark struck by the lepton.  Evidence of this phenomena has been reported by 
Scheinbein {\it et al}.~\cite{Schienbein:2009kk}

\begin{figure}[!]
 %\begin{center}
\includegraphics[width=7.5cm]{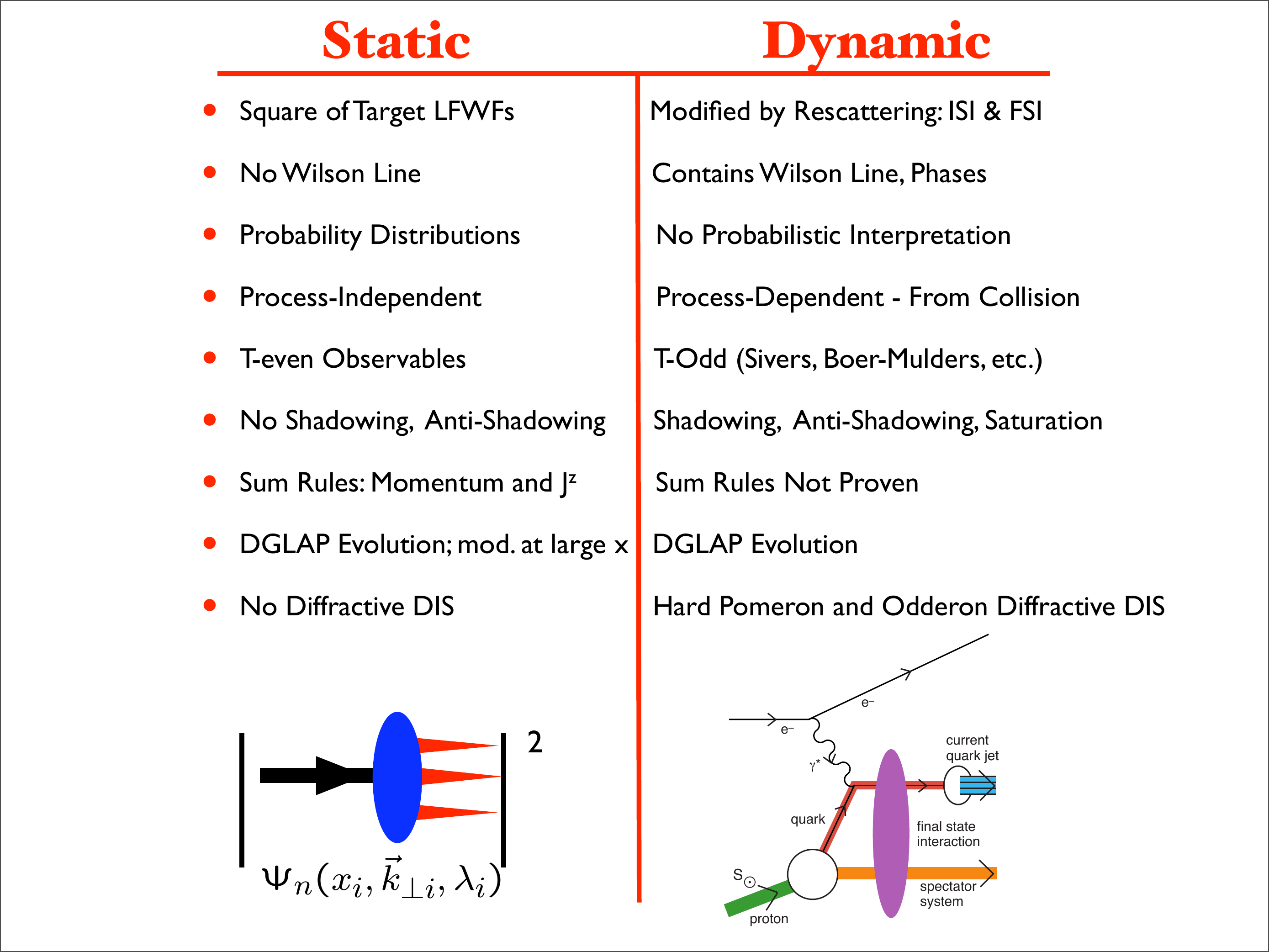}
%\end{center}
\caption{Static vs dynamic structure functions}
\label{figstatdyn}  
\end{figure}

The distinction 
between static structure functions; i.e., the probability distributions  computed from the square of the light-front wavefunctions, versus the nonuniversal dynamic structure functions measured in deep inelastic scattering is summarized in Fig. \ref{figstatdyn}.

\section{Conclusions}

Light-Front Holography is one of the most remarkable features of AdS/CFT.~\cite{Maldacena:1997re}  It  allows one to project the functional dependence of the wavefunction $\Phi(z)$ computed  in the  AdS fifth dimension to the  hadronic frame-independent light-front wavefunction $\psi(x_i, \mbf{b}_{\perp i})$ in $3+1$ physical space-time. The 
variable $z $ maps  to the LF variable $ \zeta(x_i, \mbf{b}_{\perp i})$. To prove this, we have shown that there exists a correspondence between the matrix elements of the electromagnetic current and the energy-momentum tensor of the fundamental hadronic constituents in QCD with the corresponding transition amplitudes describing the interaction of string modes in anti-de Sitter space with the external sources which propagate in the AdS interior. The agreement of the results for both electromagnetic and gravitational hadronic transition amplitudes provides an important consistency test and verification of holographic mapping from AdS to physical observables defined on the light-front.   The transverse coordinate $\zeta$ is closely related to the invariant mass squared  of the constituents in the LFWF  and its off-shellness  in  the LF kinetic energy,  and it is thus the natural variable to characterize the hadronic wavefunction.  In fact $\zeta$ is the only variable to appear in the light-front Schr\"odinger equations predicted from AdS/QCD in the limit of zero quark masses.  

The use of the invariant coordinate $\zeta$ in light-front QCD
allows the separation of the dynamics of quark and gluon binding
from the kinematics of constituent spin and internal
orbital angular momentum. The result is a single-variable LF
Schr\"odinger equation  which determines the spectrum
and  LFWFs of hadrons for general spin and
orbital angular momentum. 
This LF wave equation serves as a first approximation to QCD and is equivalent to the
equations of motion which describe the propagation of spin-$J$ modes
on  AdS~\cite{deTeramond:2008ht}.
The AdS/LF equations
correspond to the kinetic energy terms of  the partons inside a
hadron, whereas the interaction terms build confinement. Since
there are no interactions up to the confining scale in this approximation, there are no
anomalous dimensions. 
The eigenvalues of these equations for both meson and baryons give a good representation of the observed hadronic spectrum, especially in the case of the soft-wall model. The predicted LFWFs have excellent phenomenological features, including predictions for the  electromagnetic form factors and decay constants. 
This may explain the experimental success of
power-law scaling in hard exclusive reactions where there are no
indications of  the effects of anomalous dimensions.

Nonzero quark masses are naturally incorporated into the AdS/LF predictions~\cite{Brodsky:2008pg} by including them explicitly in the LF kinetic energy  $\sum_i ( {\mbf{k}^2_{\perp i} + m_i^2})/{x_i}$. Given the nonpertubative LFWFs one can predict many interesting phenomenological quantities such as heavy quark decays, generalized parton distributions and parton structure functions.

We also note the distinction between between static structure functions such as the probability distributions  computed from the square of the light-front wavefunctions versus dynamical structure functions which include the effects of rescattering.  We have also shown that the LF Hamiltonian formulation of quantum field theory provides a natural formalism to compute hadronization at the amplitude level.

The AdS/QCD model is semiclassical, and thus it only predicts the lowest valence Fock state structure of the hadron LFWF. One can systematically improve the holographic approximation by
diagonalizing the QCD LF Hamiltonian on the AdS/QCD basis, or by
generalizing the variational and other systematic methods used in
chemistry and nuclear physics~\cite{Vary:2009gt}.
The action of the non-diagonal terms
in the QCD interaction Hamiltonian also generates the form of the higher
Fock state structure of hadronic LFWFs.  In
contrast with the original AdS/CFT correspondence, the large $N_C$
limit is not required to connect LF QCD to
an effective dual gravity approximation.

\section*{Acknowledgments}
Presented by SJB at Light Cone 2009: Relativistic Hadronic And Particle Physics, 8-13 Jul 2009, S\~ao Jos\'e dos Campos, Brazil.  We are grateful to Professor Tobias Frederico  and his colleagues at the Instituto Tecnol\'ogico de Aeron\'autica (ITA) , for their outstanding hospitality.  We thank Carl Carlson, Stan Glazek, Paul Hoyer, Dae Sung Hwang, Bo-Qiang Ma, Pieter Maris, Craig Roberts, Ivan Schmidt, Robert Shrock, and James Vary  for helpful conversations and collaborations. 
Section 3 is based on collaborations with Robert Shrock.
This research was 
supported by the Department of Energy  contract DE--AC02--76SF00515.  SLAC-PUB-13778.

\end{document}